\documentclass[letterpaper,onecolumn,draftclsnofoot]{IEEEtran}
\usepackage[utf8]{inputenc} 
\usepackage[numbers,compress]{natbib}
\usepackage[cmex10]{amsmath}
\usepackage{amsfonts,mathrsfs,mathdots,mathtools,amsthm,amssymb,centernot}
\usepackage{algorithmic}
\usepackage{graphicx}
\usepackage{textcomp}
\usepackage{xcolor}
\usepackage{tikz}
\usepackage{pgfplots}
\usepackage{subcaption}
\usepackage{dsfont}
\usepackage{pifont}
\usepackage{siunitx}
\usepackage{comment}
\usepackage{pgfplotstable}
\usepackage[ruled]{algorithm2e}
\usepackage{hyperref}
\usepackage{color,soul}
\usepackage{bm}	
\usepackage[inline]{enumitem}
\usepackage{nicefrac}
\usepackage{subcaption}
\usepackage{colortbl}
\usepackage{booktabs}
\usepackage{diagbox}
\usepackage{url}
\usepackage{ifthen}
\usepackage{balance}
\usepackage{xfrac}

\DeclareMathOperator*{\argmax}{arg\,max}
\DeclareMathOperator*{\argmin}{arg\,min}

\def\supp{\textnormal{supp}}

\theoremstyle{definition}
\newtheorem{definition}{Definition}

\newtheorem{theorem}{Theorem}

\newtheorem{corollary}{Corollary}
\newtheorem{proposition}{Proposition}
\newtheorem{example}{Example}
\newtheorem{remark}{Remark}

\newcommand{\bE}{\ensuremath{\mathbb{E}}}

\newcommand{\bP}{\ensuremath{\mathbb{P}}}

\newcommand{\bR}{\ensuremath{\mathbb{R}}}

\newcommand{\cE}{\ensuremath{\mathcal{E}}}

\newcommand{\cL}{\ensuremath{\mathcal{L}}}

\newcommand{\cU}{\ensuremath{\mathcal{U}}}

\newcommand{\cW}{\ensuremath{\mathcal{W}}}
\newcommand{\cX}{\ensuremath{\mathcal{X}}}
\newcommand{\cY}{\ensuremath{\mathcal{Y}}}

\newcommand{\sfc}{\ensuremath{\mathsf{c}}}

\newcommand{\sfr}{\ensuremath{\mathsf{r}}}

\newcommand{\ind}{\ensuremath{\mathbf{1}}}

\mathtoolsset{showonlyrefs}
\allowdisplaybreaks
\begin{document}
\title{Quantifying Privacy via Information Density} 

 \author{%
   \IEEEauthorblockN{Leonhard Grosse\IEEEauthorrefmark{1},
                     Sara Saeidian\IEEEauthorrefmark{1},
                     Parastoo Sadeghi\IEEEauthorrefmark{2},
                     Tobias J. Oechtering\IEEEauthorrefmark{1},
                      and Mikael Skoglund\IEEEauthorrefmark{1}}%
                      
   \IEEEauthorblockA{\IEEEauthorrefmark{1}%
                    KTH Royal Instutite of Technology, Stockholm, Sweden,
                     \{lgrosse, saeidian, oech, skoglund\}@kth.se}
 \IEEEauthorblockA{\IEEEauthorrefmark{2}%
                     University of New South Wales,
                     Canberra, Australia,
                     p.sadeghi@unsw.edu.au}
 }

\maketitle

\begin{abstract}
We examine the relationship between privacy metrics that utilize information density to measure information leakage between a private and a disclosed random variable. Firstly, we prove that bounding the information density from above or below in turn implies a lower or upper bound on the information density, respectively. Using this result, we establish new relationships between local information privacy, asymmetric local information privacy, pointwise maximal leakage and local differential privacy. We further provide applications of these relations to privacy mechanism design. Furthermore, we provide statements showing the equivalence between a lower bound on information density and risk-averse adversaries. More specifically, we prove an equivalence between a guessing framework and a cost-function framework that result in the desired lower bound on the information density.
\end{abstract}


\section{Introduction}
\label{sec:intro}
\emph{Calculated data privacy} \cite{oechtering2023calculated} has received significant attention due to the increasing invasion of individual's privacy in modern data-intensive applications. 
To give every individual sovereignty over their own personal data without relying on a trusted data-collecting third party, the \emph{local} model of data privacy is particularly well suited for modern use-cases where sensitive data is shared from, e.g., smart phones, or other personal devices. To this end, \emph{local differential privacy} (LDP) \cite{duchi2013LDPminmaxDEF} has been shown to be a useful metric in various scenarios. 

LDP is a symmetric metric that bounds the logarithm of the ratio of any two values in the channel transition probability matrix from below and above by $-\epsilon$ and $\epsilon$, respectively. In an information-theoretic context, recently many alternatives to this approach have been proposed \cite{10352344,asoodeh2018estimation,rassouli2021perfect,rassouli2019TVDprivacy,9360888,asoodeh2014notes,asoodeh2015maximal,asoodeh2016information,liao2017hypothesis,makhdoumi2014information,zamani2023privacy,du2017principal,wang2019privacy}, which have been applied to privacy mechanism design in, e.g., \cite{rassouli2019TVDprivacy,lopuhaa2020privacy,8849440,sadeghi2021properties,9162276,saeidian2021hamming,zamani2021design,zamani2021data}.
A methodology similar to the LDP framework has been applied to define \emph{local information privacy} (LIP) \cite{6483382,jiang2018context,jiang2020LIP,jiang2021LIPcontextaware}. More specifically, LIP bounds the information density between private and public realizations from below and above by $-\epsilon$ and $\epsilon$, respectively. As a ratio of posterior to prior distributions, LIP therefore is a context-aware (prior-dependent) notion of privacy, and hence not a symmetric definition. As shown in \cite{zarrabian2023lift,9965910}, enforcing symmetric bounds on this non-symmetric measure can cause a significant degradation in utility, thereby erasing the potential benefit of a context-aware privacy notion. To remedy this issue, \emph{asymmetric local information privacy} (ALIP) \cite{zarrabian2023lift,9965910} generalizes LIP by imposing different lower and upper bounds on the information density values, allowing for an additional degree of freedom in trading off utility against privacy compared to symmetric LIP.

Nonetheless, the bounds in (A)LIP are applied axiomatically, and not derived with a specific operational motivation in mind. While the upper bound in (A)LIP has later been shown to follow from such operational models \cite{saeidian2023pointwise,saeidian2023pointwisegeneral}, the meaning of the lower bound remains largely unmotivated. In contrast, the formulation of \emph{pointwise maximal leakage} (PML) is done starting from a robust operational adversarial attack formulation in \cite{saeidian2023pointwise}. For finite alphabets, the authors then show that an upper bound on information density provides a privacy guarantee with respect to such attacks.\footnote{More generally, PML is defined as an extension of \emph{g-leakage} \cite{ACPS12milu} to arbitrary probability spaces and is not restricted to discrete random variables.} That is, ALIP and PML---although derived using two very different approaches---result in a similar way of measuring privacy in the discrete case. However, as briefly pointed out in, e.g., \cite{saeidian2023pointwise}, it is clear that the information density values between each realization of private and public data cannot be bounded completely independently form each other: The probabilities in the ratios have to satisfy the basic conditions of valid probability distributions; these stochasticity requirements impose some restriction on how the lower and upper bound on information density behave in relation to each other. It is therefore reasonable to assume that privacy measures based on bounding information density are closely related. In this paper, we will make the relation between these measures explicit and move towards a unification of the privacy frameworks based on bounding information density. This unification enables a direct transfer of results from PML to (A)LIP and vice versa. Further, by discussing and extending the operationalization of the lower bound in (A)LIP, we provide deeper insights into the privacy guarantees based on the considered measures.
\subsection{Summary of contributions and outline}
In Section \ref{sec:liftasym}, we present new relations between PML, (A)LIP and LDP using bounds on information density derived from simple row-stochasticity conditions. We show how these relations can be used to design optimal privacy mechanisms in Section \ref{sec:optmech}. Further, we discuss operationalizations of the lower bound on information density in Section \ref{sec:operational} alongside its consequences on privacy risk assessment.

\section{System Model and Privacy Metrics}
We represent the private data by the random variable $X\sim P_X$ defined on an alphabet $\mathcal X$.\footnote{In this paper, we assume all alphabets to be finite.} Our goal is then to release a privatized version of a realization $X = x$, denoted as the realization $y$ of the random variable $Y$. We relate $Y$ to $X$ by the conditional distribution $P_{Y|X}$ (also known as \emph{(privacy) mechanism} or \emph{kernel}). In the present discrete alphabet case and by letting $|\mathcal X|\coloneqq N$, $|\mathcal Y| \coloneqq M$, this mechanism takes the form of an $N\times M$ row-stochastic matrix, where $(P_{Y|X})_{ij} = P_{Y|X=x_i}(y_j)$ for $i \in [N]$, $j \in [M]$.\footnote{$[N] \coloneqq \{1,\dots,N\}$ denotes the set of positive integers up to $N$.} We denote the minimum probability of the distribution $P_X$ as $p_{\min} \coloneqq \min_{x \in \mathcal X}P_X(x)$.


 \subsubsection{Context-aware privacy notions}
Given private data with a prior distribution $P_X$ and mechanism $P_{Y|X}$, the \emph{information density} between outcomes $x \in \mathcal X$ and $y \in \mathcal Y$ is given by
\begin{equation}
    i(x;y) = \log \frac{P_{X|Y=y}(x)}{P_X(x)}.
\end{equation}
Based on this quantity, we consider the following context-aware (i.e., prior-dependent) privacy measures.
\begin{definition}[\emph{Local information privacy} (LIP), {{\cite{jiang2018context,6483382}}}]
    A mechanism $P_{Y|X}$ satisfies \emph{$\epsilon$-LIP} if
    \begin{equation}
        -\epsilon \leq i(x;y) \leq \epsilon \quad \forall x \in \mathcal X, \; \forall y \in \mathcal Y.
    \end{equation}
\end{definition}
It is observed in \cite{zarrabian2023lift} that 
the probability distribution of $i(x;y)$ exhibits an asymmetry in the sense that it has a skewed long tail for small negative values, and a much shorter range with a sharp fall-off for large positive values. This asymmetry, which the authors label the \emph{lift-asymmetry}, motivates their definition of \emph{asymmetric local information privacy} (ALIP): 
\begin{definition}[\emph{Asymmetric local information privacy} (ALIP), \cite{zarrabian2023lift}]
    A mechanism $P_{Y|X}$ satisfies \emph{$(\epsilon_l,\epsilon_u)$-ALIP} if
    \begin{equation}
    \label{eq:defALIPbounds}
        -\epsilon_l \leq \min_{x \in \mathcal X}i(x;y) \leq \max_{x \in \mathcal X}i(x;y) \leq \epsilon_u \quad \forall y \in \mathcal Y.
    \end{equation}
\end{definition}

Another privacy measure based on a bound on information density that is central to this work is \emph{pointwise maximal leakage} (PML) \cite{saeidian2023pointwise}. 
In contrast to LIP (and ALIP), PML is derived from threat model formulations, and therefore provides an operational definition of information leakage. In the discrete alphabet case, the maximum information density between the private variable $X$ and the considered outcome $y$ characterizes the pointwise maximal leakage from $X$ to $y$.
\begin{theorem}[\emph{Pointwise maximal leakage} (PML), {{\cite[Thm. 1]{saeidian2023pointwise}}}]
    For discret alphabets $\mathcal X$ and $\mathcal Y$, the \emph{pointwise maximal leakage} between the variable $X$ and an outcome $y \in \mathcal Y$ can be expressed as
    \begin{equation}
        \ell(X \to y) = \max_{x \in \supp(P_X)} i(x;y).
    \end{equation}
\end{theorem}
A stringent privacy guarantee using PML can be defined by bounding $\ell(X\to y)$ by a value $\epsilon$ for all $y$ in $\mathcal Y$.\footnote{Note that other guarantees can be defined, e.g., approximate or average-case guarantees. For a detailed discussion, see \cite{saeidian2023pointwise}.}
\begin{definition}[\emph{$\epsilon$-PML}, \cite{saeidian2023pointwise}]
    Assume $X$ is distributed according to $P_X$. A mechanism $P_{Y|X}$ satisfies \emph{$\epsilon$-PML} if 
    \begin{equation}
        \ell(X \to y) \leq \epsilon \quad \forall y \in \mathcal Y.
    \end{equation}
\end{definition}

In \cite{grosse2023extremal}, the authors define the notion of \emph{PML privacy regions}, which partition the space of privacy parameter $\epsilon \in [0,\infty)$ into disjoint subsets in which privacy mechanisms display similar disclosure behavior in terms of inference limits (see \cite{saeidian2023inferential}). These regions specify upper bounds on the number of elements in the mechanisms matrix that can become zero for a mechanism satisfying $\epsilon$-PML (see \cite[Lemma 2]{grosse2023extremal}). In the context of local information privacy, the region in which $\epsilon < \log\frac{1}{1-p_{\min}}$ (called the \emph{high-privacy regime}) is particularly interesting. Any mechanism in the high-privacy regime guarantees \emph{absolute disclosure prevention} \cite{saeidian2023inferential}, and therefore does not allow for \emph{any} zero probability assignments in the mechanism matrix. As a result, mechanisms satisfying $\epsilon$-PML in the high-privacy regime also provide some $\epsilon_{\text{LIP}}$-LIP and some $\epsilon_{\text{LDP}}$-LDP guarantee. In this region, we can therefore derive meaningful relationships providing privacy guarantees with respect to other privacy measures.

\subsubsection{Local differential privacy}
We further relate the above privacy measures to local differential privacy.
\begin{definition}[\emph{Local differential privacy} (LDP), \cite{duchi2013LDPminmaxDEF}]
    A mechanism $P_{Y|X}$ satisfies \emph{$\epsilon$-LDP}, if for all  $x,x' \in \mathcal X$ and for all $y \in \mathcal Y$ we have
    \begin{equation}
        \frac{P_{Y|X=x'}(y)}{P_{Y|X=x}(y)} \leq e^{\epsilon}.
    \end{equation}
\end{definition}
Note that the definition of LDP does not assume a specific prior distribution of the private data $X$. A privacy guarantee with LDP therefore holds true regardless of an adversary's prior knowledge about the distribution $P_X$. 

\subsubsection{New and existing relationships}
\begin{table}[!t]
  \centering
  \begin{tabular}{c|c|c|c|c}
  \toprule
    \diagbox{from}{to} & $\epsilon$-PML & $\epsilon$-LIP & $(\epsilon_l,\epsilon_u)$-ALIP & $\epsilon$-LDP\\
    \midrule
    $\epsilon$-PML & - & \cellcolor{green!25}Cor. \ref{corr:PML->LIP} & \cellcolor{green!25} Cor. \ref{corr:PML->ALIP} & \cellcolor{green!25} Cor. \ref{corr:PML->LDP} \\

    $\epsilon$-LIP & \cellcolor{yellow!25}\cite{saeidian2023pointwise} & - & \cellcolor{yellow!25}\cite{zarrabian2023lift} & \cellcolor{yellow!25}\cite[Thrm. 1]{jiang2021LIPcontextaware} \\
  
    $(\epsilon_l,\epsilon_u)$-ALIP & \cellcolor{green!25} Prop. \ref{prop:liftasym}, \ref{prop:liftasym2}\ & \cellcolor{yellow!25}\cite{zarrabian2023lift} & - & \cellcolor{yellow!25}\cite{zarrabian2023lift} \\
 
    $\epsilon$-LDP & \cellcolor{yellow!25}\cite{saeidian2023pointwise} & \cellcolor{yellow!25}\cite{jiang2021LIPcontextaware} & \cellcolor{yellow!25}\cite{zarrabian2023lift} & - 
    \\\bottomrule

  \end{tabular}
  \caption{Relationship between local privacy metrics based on bounding information density. New results in this paper are marked in green, existing relationship are colored in yellow.}
  \label{tab:mytable}
\end{table}
The relations between the above measures are summarized in Table \ref{tab:mytable}. New relations derived in this paper are highlighted against previously known relations. In Section \ref{sec:newrelations}, we will provide proofs for the new relationships between measures (highlighted in green). The proofs of the previously known relations (highlighted in yellow) can be found in the cited literature.

\section{Bounding Information Density for Privacy}
\label{sec:liftasym}
In the following statements, we esablish the connection between upper and lower bounds on information density.

\begin{proposition}
\label{prop:liftasym}
    Assume $X$ is distributed according to $P_X$. For any $y \in \mathcal Y$ and $\epsilon_u < \log \frac{1}{1-p_{\min}}$, if $\max_{x} i(x;y) \leq \epsilon_u$, then $\min_{x} i(x;y) \geq -\epsilon_l(\epsilon_u)$, where
    \begin{equation}
    \label{eq:epsl(epsu)}
        \epsilon_l(\epsilon_u) = \log\frac{p_{\min}}{1-e^{\epsilon_u}(1-p_{\min})}.
    \end{equation} 
\end{proposition}
\begin{IEEEproof}
       Fix some $y \in \mathcal Y$ and $i \in [N]$. Then we have 
    \begin{equation}
        P_{X|Y=y}(x_i) + \sum_{j \in [N]\setminus\{i\}} P_{X|Y=y}(x_j) = 1.
    \end{equation}
    Now, assume $i(x;y) \leq \epsilon_u$. We then have $P_{X|Y=y}(x) \leq e^{\epsilon_u}P_X(x)$ and therefore we get the upper bound
    \begin{equation}
        \sum_{x \in \mathcal X}P_{X|Y=y}(x) \leq P_{X|Y=y}(x_i) + \sum_{j \in [N]\setminus\{i\}} e^{\epsilon_u}P_X(x_j),
    \end{equation}
    implying that
    \begin{equation}
        P_{X|Y=y}(x_i) \geq 1-e^{\epsilon_u}(1-P_X(x_i)).
    \end{equation}
    From this, we can lower bound the minimum information density value as 
    \begin{equation}
             \label{eq:liftlimit}
             \min_{x \in \mathcal X}\frac{P_{X|Y=y}(x)}{P_X(x)} \geq \min_{x\in \mathcal X}\frac{1-e^{\epsilon_u}(1-P_X(x))}{P_X(x)} = \frac{1-e^{\epsilon_u}(1-p_{\min})}{p_{\min}}.
    \end{equation}
    This holds for all $\epsilon_u \geq 0$ and all $y \in \mathcal Y$. Solving \eqref{eq:liftlimit} for $\epsilon_l$ results in the desired lower bound. 
\end{IEEEproof}
Next, we prove the reverse of Proposition \ref{prop:liftasym}, that is, we prove that any lower bound on the minimum information density implies an upper bound on its maximum.
\begin{proposition}
\label{prop:liftasym2}
    For any $\epsilon_l \geq 0$ and any $y \in \mathcal Y$, $\min_x i(x;y) \geq -\epsilon_l$ implies $\max_x i(x;y) \leq \epsilon_u(\epsilon_l)$, where 
    \begin{equation}
    \epsilon_u(\epsilon_l) = \log\biggl(\frac{1-e^{-\epsilon_l}(1-p_{\min})}{p_{\min}}\biggr).
    \end{equation}
\end{proposition}
\begin{IEEEproof}
        Assume the mechanism $P_{Y|X}$ satisfies $\min_x i(x;y) \geq e^{-\epsilon_l}$ for some $y \in \mathcal Y$. Then we also have
        \begin{equation}
            e^{-\epsilon_l} \leq \frac{P_{X|Y=y}(x)}{P_X(x)} \quad \forall x \in \mathcal{X}.
        \end{equation}
        Fix some $i \in [N]$. From the lower bound above we get 
        \begin{equation}
            1 = \sum_{x \in \mathcal X}\frac{P_{X|Y=y}(x)}{P_X(x)}P_X(x) \geq P_{X|Y=y}(x_i) + \sum_{j \in [N] \setminus \{i\}}e^{-\epsilon_l}P_X(x_j) = P_{X|Y=y}(x_i) + e^{-\epsilon_l}(1-P_X(x_i))).
        \end{equation}
      We therefore get the upper bound
        \begin{equation}
            P_{X|Y=y}(x_i) \leq 1-e^{-\epsilon_l}(1-P_X(x_i))
        \end{equation}
        and renormalizing yields
        \begin{align}
            \frac{P_{X|Y=y}(x_i)}{P_X(x_i)} \leq \frac{1-e^{-\epsilon_l}(1-P_X(x_i)}{P_X(x_i)}.
        \end{align}
        Since $e^{-\epsilon_l} \leq 1$, $\frac{1-e^{-\epsilon_l}(1-P_X(x))}{P_X(x)}$ is non-increasing in $P_X(x)$, we get the upper bound
        \begin{equation}
            \max_{x \in \mathcal X} i(x;y) \leq \frac{1-e^{-\epsilon_l}(1-p_{\min})}{p_{\min}}. \IEEEQEDhereeqn
     \end{equation} 
    \end{IEEEproof}

\begin{figure}[!t]
    \begin{subfigure}[b]{.49\textwidth}
    \hspace{-.5cm}
        \includegraphics[scale=0.4]{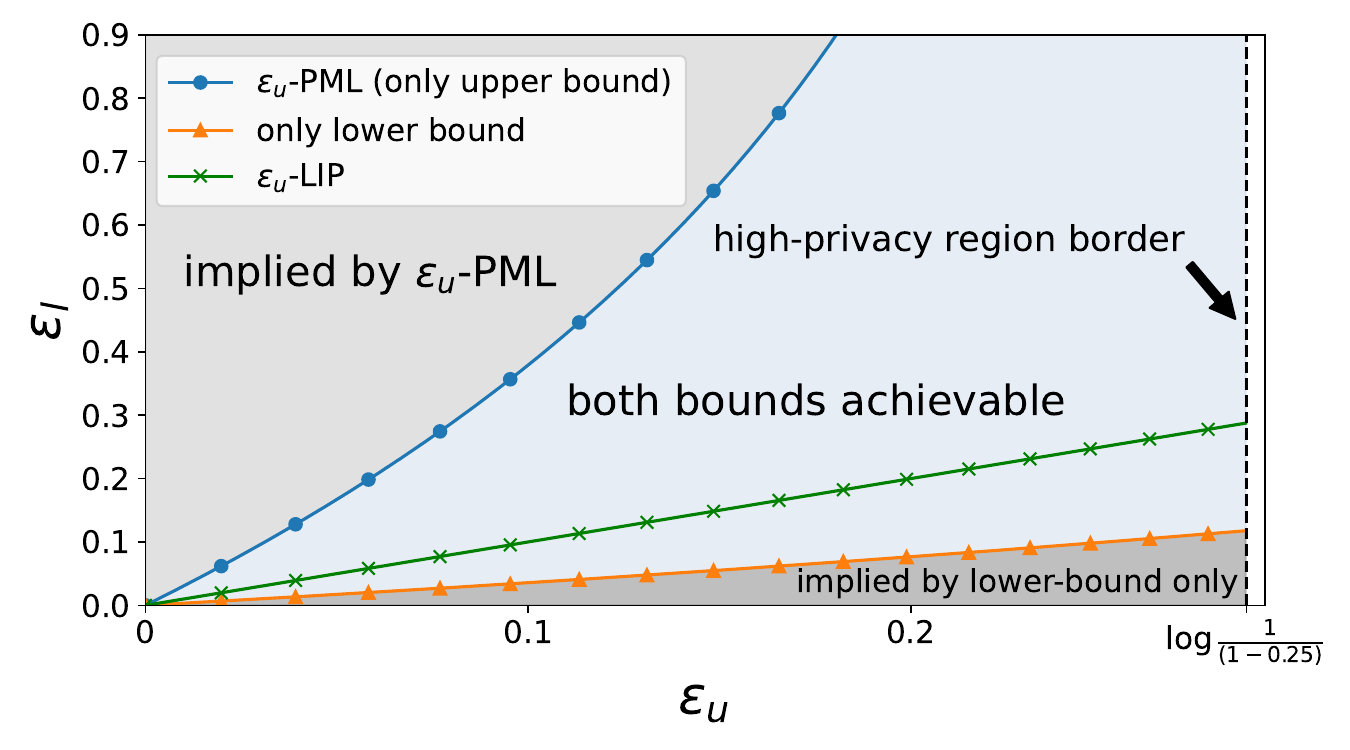}
        \caption{$N=4$, $P_X = (\nicefrac{1}{4},\nicefrac{1}{4},\nicefrac{1}{4},\nicefrac{1}{4})^T$}
        \label{fig:eps_u-vs-eps_l:subfig:N4}
    \end{subfigure}
    \begin{subfigure}[b]{.49\textwidth}
        \includegraphics[scale=0.4]{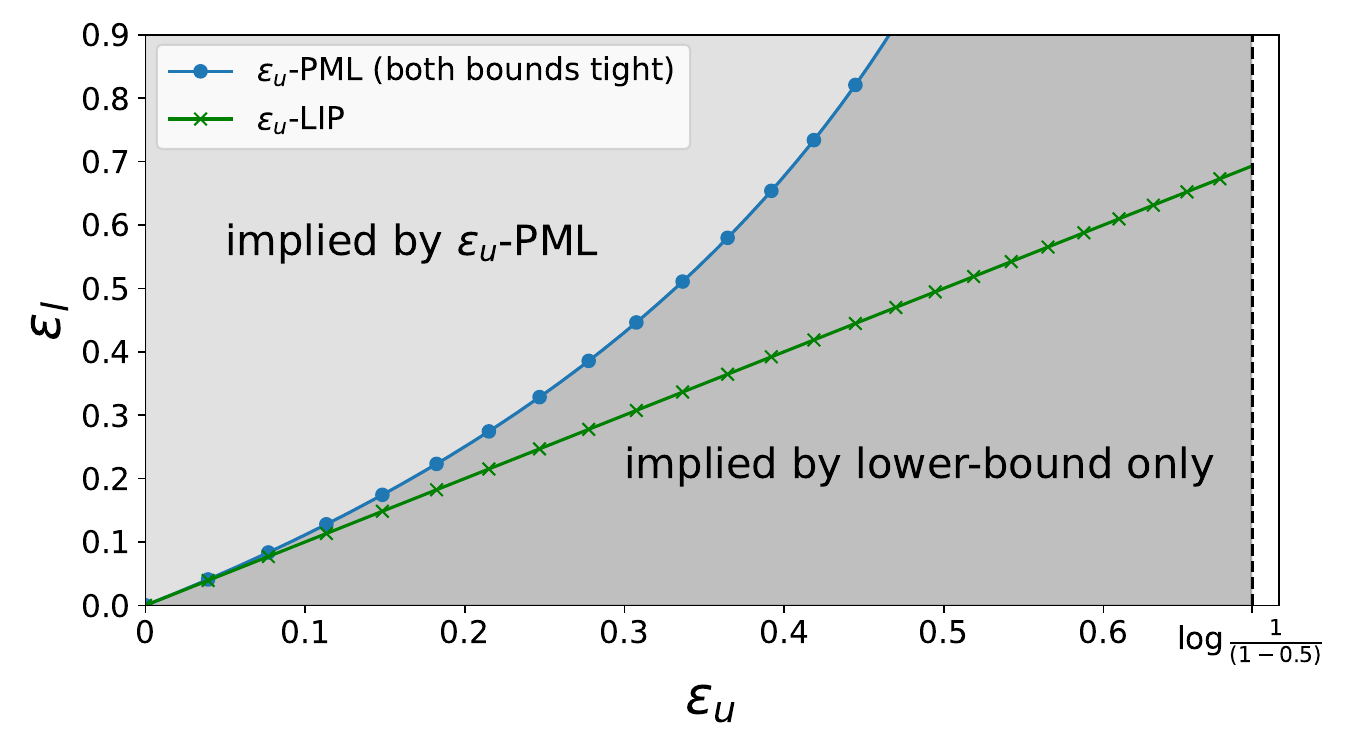}
        \caption{$N=2$, $P_X = (\nicefrac{1}{2},\nicefrac{1}{2})^T$}
        \label{fig:eps_u-vs-eps_l:subfig:N2}
    \end{subfigure}
\centering
    \caption{Relationship between the parameters $\epsilon_l$ and $\epsilon_u$ for a uniform prior distribution and $N\in \{2,4\}$. PML offers a tight bound on $\epsilon_u$, which yields a corresponding AILP guarantee. Lower-bounding the information density results in a corresponding upper bound. The region in between these two bounds represents values in which both upper and lower bound can be achieved simultaneously. In the binary case, upper and lower bound coincide.}  
\label{fig:eps_u-vs-eps_l}
\end{figure}  

Figure \ref{fig:eps_u-vs-eps_l} illustrates the relation between the upper and lower bound in the ALIP framework. We see that in the high-privacy regime, both the upper and lower bounds on information density \emph{alone} impose bounds in the opposite direction. Hence, in certain ranges of values for $(\epsilon_u,\epsilon_l)$, either bounding $\min_x i(x;y)$ or bounding $\max_x i(x;y)$ is superfluous. In this case, constraining either the maximum or the minimum of the information density implicitly imposes a stricter requirement on the minimum or the maximum information density than the bound specified by $\epsilon_l$ or $\epsilon_u$, respectively. As a result, we have two regions (shaded grey in Figure \ref{fig:eps_u-vs-eps_l}) where one constraint is always redundant. In the region between these two cases, privacy mechanisms might achieve both bounds at the same time. In the depicted case, symmetric local information privacy falls into this region. Interestingly, for the case of uniform priors and $N=2$, it can be shown that the region in which both bounds can be achieved reduces to the line given by $\epsilon_u$-PML. That is, an upper bound for the information density automatically implies its tight lower bound, and vice versa. This relationship will be discussed more in Example \ref{ex:binaryuniform}.

\label{sec:newrelations}
We utilize the above insight to derive new relations between privacy metrics in the following.
\subsection{ALIP guarantees implied by PML}
It is shown in \cite{saeidian2023pointwise} that $\epsilon$-LIP implies $\epsilon$-PML. However, using the privacy regions introduced in \cite{grosse2023extremal}, it is easy to see that in the \emph{high-privacy}-regime, $\epsilon$-PML necessarily also implies  $(\epsilon',\epsilon)$-ALIP with some $\epsilon' < \infty$.\footnote{This can easily be shown by contradiction, as $\epsilon' \nless \infty$ implies some zero information density value, which cannot exist in the high-privacy regime.} The following statement provides an explicit formulation.

\begin{corollary}
\label{corr:PML->ALIP}
    For any $\epsilon \leq \log \frac{1}{1-p_{\min}}$, $\epsilon$-PML implies $(\epsilon_l(\epsilon),\epsilon)$-ALIP, with $\epsilon_l(\epsilon)$ according to \eqref{eq:epsl(epsu)}.
\end{corollary}
\begin{IEEEproof}
Follows directly from Proposition \ref{prop:liftasym}.
    \end{IEEEproof}
    Note that for any $\epsilon \geq -\log(1-p_{\min})$, the value of $\epsilon_l$ becomes unbounded. Therefore, $\epsilon$-PML only implies \emph{useful} ALIP guarantees in the high-privacy regime $\epsilon < \log\frac{1}{1-p_{\min}}$. Further, the optimal PML mechanism in the high-privacy regime presented in \cite{grosse2023extremal} achieves the bound in Corollary \ref{corr:PML->ALIP} for some $y$.

    \subsection{LIP guarantees implied by PML}
    Local information privacy can be obtained from ALIP by setting $\epsilon_u = \epsilon_l$. Obviously, $\epsilon$-LIP implies $\epsilon$-PML. An implication in the other direction can be characterized by the relation in \eqref{eq:epsl(epsu)}.
    \begin{corollary}
    \label{corr:PML->LIP}
        For any $\epsilon < \log\frac{1}{1-p_{\min}}$, $\epsilon$-PML implies $\epsilon_{\text{LIP}}$-LIP, where
        \begin{equation}
            \epsilon_{\text{LIP}} \coloneqq \log\frac{p_{\min}}{1-e^{\epsilon}(1-p_{\min})}.
        \end{equation}
    \end{corollary}
    \begin{IEEEproof}
        Follows directly from Corollary \ref{corr:PML->ALIP} by noting that $(\epsilon_l,\epsilon_u)$-ALIP implies $\max\{\epsilon_l,\epsilon_u\}$-LIP.
    \end{IEEEproof}

\subsection{LDP guarantees implied by PML}
It is shown in \cite{jiang2021LIPcontextaware} that $\epsilon$-LIP implies $2\epsilon$-LDP. This result is extended in \cite{zarrabian2023lift} for ALIP, stating that $(\epsilon_l,\epsilon_u)$-ALIP implies $(\epsilon_l + \epsilon_u)$-LDP. Using the results for PML and ALIP in Corollary \ref{corr:PML->ALIP}, we can obtain an implication from PML to LDP.

\begin{corollary}
\label{corr:PML->LDP}
    For any $\epsilon < \log \frac{1}{1-p_{\min}}$, $\epsilon$-PML implies $(\epsilon_l(\epsilon) + \epsilon)$-LDP, with $\epsilon_l(\epsilon)$ according to \eqref{eq:epsl(epsu)}.
\end{corollary}

We conclude this section by giving the following example.
\begin{example}
\label{ex:binaryuniform}
    Assume $\mathcal X \coloneqq \{x_1,x_2\}$ and $P_X(x_1) = P_X(x_2)$. Then it is easy to check that 
    \begin{align}
    \epsilon\text{-PML} \iff \epsilon_{\text{LIP}}\text{-LIP} \iff (\epsilon_l(\epsilon),\epsilon)\text{-ALIP} \iff (\epsilon_l(\epsilon)+\epsilon)\text{-LDP}.
    \end{align}
    This is confirmed by the fact that, using the transformation $\epsilon_{\text{LDP}} \coloneqq \epsilon_l(\epsilon)+\epsilon$, the binary randomized response mechanism for $\epsilon_{\text{LDP}}$ in \cite{extremalmechanismLong} is equivalent to the optimal binary mechanism for $\epsilon$-PML and uniform priors presented in \cite{grosse2023extremal} (an instance of the high-privacy mechanism in Corollary \ref{corr:optMech}). 
\end{example}

\section{Application to Mechanism Design}
\label{sec:optmech}
From the above derivations, we can see that there are $(\epsilon_l,\epsilon_u)$-ALIP guarantees implied by $\epsilon_u$-PML. As a result, for cases in which the lower bound on information density is implied by the upper bound, that is, for $\epsilon_l \geq \epsilon_l(\epsilon_u)$, mechanisms optimal for $\epsilon$-PML are also optimal for $(\epsilon_l,\epsilon_u)$-ALIP, as is shown in the following proposition.

\begin{proposition}
\label{prop:optimalPML}
 Assume $X\sim P_X$. Then, for all $\epsilon_u < \frac{1}{1-p_{\min}}$, the optimal $(\epsilon_l,\epsilon_u)$-ALIP mechanism with $\epsilon_l \geq \epsilon_l(\epsilon_u)$ is identical to the mechanism optimal for $\epsilon_u$-PML. 
\end{proposition}
\begin{IEEEproof}
    The proof is by contradiction. Assume utility is measured by some function $U: \mathcal{X}\times \mathcal Y \to \mathbb R$. Fix some $\epsilon_u\geq 0$ and some $\epsilon_l \geq \epsilon_l(\epsilon_u)$. Further, denote a mechanism satisfying $\epsilon_u$-PML and optimal with respect to $U$ as $P_{Y|X}^{*}$. Note that this mechanism also satisfies $(\epsilon_l,\epsilon_u)$-ALIP according to Corollary \ref{corr:PML->ALIP}. Now, assume there exists a mechanism $Q_{Y|X}^*$ satisfying $(\epsilon_l,\epsilon_u)$-ALIP such that $U(Q_{Y|X}^*) > U(P_{Y|X}^{*})$. Then, since $(\epsilon_l,\epsilon_u)$-ALIP implies $\epsilon_u$-PML, $P_{Y|X}^{*}$ could not be the optimal PML mechanism, which yields a contradiction, as desired.
\end{IEEEproof}

In \cite{grosse2023extremal}, the authors present closed-form privacy mechanisms for $\epsilon$-PML that are optimal when considering the broad class of sub-convex utility functions \cite[Definition 5]{grosse2023extremal} in the high-privacy regime. As a result of Proposition \ref{prop:optimalPML}, these mechanisms are also optimal for some ALIP guarantees in the same setup.

\begin{corollary}
\label{corr:optMech}
    Assume $X$ is distributed according to $P_X$ with full support on alphabet $\mathcal X$ of size $N$. Assume further that $\epsilon_u < \frac{1}{1-p_{\min}}$ and $\epsilon_l > \log\frac{p_{\min}}{1-e^{\epsilon_u}(1-p_{\min})}$. Then the mechanism maximizing any sub-convex utility function while satisfying $(\epsilon_l,\epsilon_u)$-ALIP is given by
    \begin{equation}
    \label{eq:optPMLsubconvex}
        P_{Y|X=x_i}^*(y_j) = \begin{cases}
           1 - e^{\epsilon_u}(1-P_X(x_i)) & i=j \\
            e^{\epsilon_u}P_X(x_j) & i \neq j
        \end{cases},
    \end{equation}
    where $i,j \in [N]$.
\end{corollary}
\begin{IEEEproof}
    Follows from Proposition \ref{prop:optimalPML} and \cite[Thm. 3]{grosse2023extremal}.
\end{IEEEproof}

\section{Operationalazing the Lower Bound on Information Density}
\label{sec:operational}
In contrast to $\max_x i(x;y)$, which finds an operational meaning in the PML framework and is extensively discussed in \cite{saeidian2023pointwise, saeidian2023inferential}, $\min_x i(x;y)$ remains a significantly under-discussed quantity. This section explores the implications of achieving privacy by imposing a lower bound on $\min_x i(x;y)$. More specifically, we demonstrate that $\min_x i(x;y)$ describes the amount of information leaking to \emph{risk-averse} adversaries, as opposed to the \emph{opportunistic} adversaries of PML. The notions of information leakage introduced in this section are closely related to \emph{maximal cost leakage} and \emph{maximal realizable cost} defined in \cite{IssaMaxL}, and we discuss their connections in Section~\ref{ssec:lower_bound_discuss}. 

\subsection{Quantifying Information Leaked to Risk-averse Adversaries}
In~\cite{saeidian2023pointwise}, PML is defined by analyzing two (seemingly) distinct threat models: the \emph{randomized function} model of leakage, positing that adversaries seek to guess the values of randomized functions of $X$, and the \emph{gain function} model, assuming that adversaries try to maximize arbitrary non-negative gain functions. Moreover, it has been established in~\cite[Thm. 2]{saeidian2023pointwise} that these two threat models are equivalent. Our discussions in this section regarding $\min_x i(x;y)$ parallel those in the context of PML. 

First, suppose an adversary attempts to guess the value of a randomized function of $X$, denoted by $U$. Upon observing $y \in \cY$, the adversary formulates a guess of $U$, denoted by $\hat U$. The guessing strategy $P_{\hat U \mid Y}$ is optimally selected to minimize the probability of an incorrect guess. To quantify the risk incurred by such an adversary, we compare the adversary's probability of incorrectly guessing $U$ having access to $y$ and the probability of incorrectly guessing $U$ without access. Consequently, we define 
\begin{equation}
\label{eq:U_def}
    \Lambda_{U}(X \to y) \coloneqq \log \; \frac{\inf_{P_{\hat U}}\; \bP [U \neq \hat{U}]}{\inf_{P_{\hat U \mid Y}} \; \bP [U \neq \hat{U} \mid Y=y]}. 
\end{equation}

In general, we may not know what function of $X$ the adversary is interested in. Therefore, to achieve robustness, we define 
\begin{align}
\label{eq:lambda_sup_U}
    \Lambda_{P_{XY}}(X \to y) &\coloneqq \sup_{U: U-X-Y} \; \Lambda_{U}(X \to y) \\
    &=\sup_{U: U-X-Y} \; \log \; \frac{\inf_{P_{\hat U}}\; \bP [U \neq \hat{U}]}{\inf_{P_{\hat U \mid Y}} \; \bP [U \neq \hat{U} \mid Y=y]},
\end{align}
that is, we maximize $\Lambda_{U}(X \to y)$ over all possible $U$'s satisfying the Markov chain $U-X-Y$. 

Below, we show how $\Lambda_{P_{XY}}(X \to y)$ relates to $\min_x i(x;y)$.
\begin{theorem}
\label{thrm:operationalization}
    Assume $X$ is distributed according to $P_X$. Then for all $y \in \mathcal Y$, lower bounding $\min_{x \in \cX} i(x;y)$ is equivalent to upper bounding $\Lambda_{P_{XY}}(X \to y)$, that is,
    \begin{equation}
    -\epsilon \leq \min_{x \in \mathcal X}i(x;y) \iff \Lambda_{P_{XY}}(X \to y) \leq \epsilon.
    \end{equation}
\end{theorem}
\begin{IEEEproof}
We prove this statement by showing that $\Lambda_{P_{XY}}(X \to y)$ is both upper bounded and lower bounded by $\max_x (-i(x;y))$.
\subsubsection{Upper bound}
Let $u_1 \in \argmax P_U(u)$ and $u_2 \in \argmax P_{U \mid Y=y}(u)$. We have 
\begin{align*}
    \exp\big(\Lambda_{P_{XY}}(X \to y)\big) &= \sup_{U: U-X-Y} \; \frac{1 - \max_u P_U(u)}{1 - \max_u P_{U \mid Y=y}(u)}\\
    &= \sup_{U: U-X-Y} \; \frac{\sum_{u \neq u_1} P_U(u)}{\sum_{u \neq u_2} P_{U \mid Y=y}(u)}. 
\end{align*}

Since $P_U(u_1) = \max_u P_{U}(u) \geq P_U(u_2)$, we get  
\begin{align*}
    \frac{\sum_{u \neq u_1} P_U(u)}{\sum_{u \neq u_2} P_{U \mid Y=y}(u)} &\leq \frac{\sum_{u \neq u_2} P_U(u)}{\sum_{u \neq u_2} P_{U \mid Y=y}(u)}\\
    &\leq \max_{u \neq u_2} \frac{P_{U}(u)}{P_{U \mid Y=y}(u)}\\
    &= \max_{u \neq u_2} \frac{\sum\limits_x P_{U \mid X=x}(u) P_X(x)}{\sum\limits_x P_{U \mid X=x}(u) P_{X \mid Y=y}(x)}\\
    &\leq \max_{x} \frac{P_{X}(x)}{P_{X \mid Y=y}(x)}. 
\end{align*}
We conclude that $\exp\big(\Lambda_{P_{XY}}(X \to y)\big) \leq \max_x \; \frac{P_X(x)}{P_{X \mid Y=y}(x)}$. 

\subsubsection{Lower bound}

Let $x^* \in \argmax_x \frac{P_X(x)}{P_{X \mid Y=y}(x)}$, and define $V \coloneqq \ind_{X \setminus \{x^*\}}$. Then, $V$ is a binary random variable with $P_{V}(0) = 1 - P_{V}(1) = P_X(x^*)$.

Let $k$ be a large integer. We define a random variable $W$ with alphabet $\cW = \{1, \ldots, k+1\}$ and induced by the kernel 
\begin{equation*}
    P_{W \mid V=0}(w) = \begin{cases}
    \frac{1}{k} & \text{if} \; w \in \{1, \ldots, k\},\\
    0 & \text{if} \; w = k+1,\\
    \end{cases}
\end{equation*}
and 
\begin{equation*}
    P_{W \mid V=1}(w) = \begin{cases}
    0 & \text{if} \; w \in \{1, \ldots, k\},\\
    1 & \text{if} \; w = k+1.\\
    \end{cases}
\end{equation*}
Then, $P_W(w) = \frac{P_{V}(0)}{k} = \frac{P_X(x^*)}{k}$ for $w \in [k]$ and $P_W(k+1) = P_{V}(1)$. Thus, taking $k$ to be sufficiently large, we can ensure that $P_W(k+1) = \max_w P_W(w)$. Also, note that since the Markov chain $W-V-X-Y$ holds, then $P_{W \mid Y=y}(w) = \frac{P_{V \mid Y=y} (0)}{k} = \frac{P_{X \mid Y=y}(x^*)}{k}$ for $w \in [k]$ and $P_{W \mid Y=y}(k+1) = P_{V \mid Y=y}(1) = P_{X \mid Y=y}(\cX \setminus \{x^*\})$. Once again, taking $k$ to be sufficiently large we can ensure that $P_{W \mid Y=y}(k+1) = \max_w P_{W \mid Y=y}(y)$. 

Since $W - V - X - Y$ is a Markov chain we get 
\begin{align}
    \exp\big(\Lambda_{P_{XY}}(X \to y)\big) &= \sup_{U: U-X-Y} \; \frac{1 - \max_u P_U(u)}{1 - \max_u P_{U \mid Y=y}(u)}\\
    &\geq \frac{1 - \max_w P_W(w)}{1 - \max_w P_{W \mid Y=y}(w)}\\
    &= \frac{\sum_{w \neq k+1} P_W(w)}{\sum_{w \neq k+1} P_{W \mid Y=y}(w)}\\
    &= \frac{P_X(x^*)}{P_{X \mid Y=y}(x^*)}\\
    &= \max_x \frac{P_X(x)}{P_{X \mid Y=y} (x)},
\end{align}
as desired.
\end{IEEEproof}

Now, suppose an adversary tries to minimize the expected value of a non-negative \emph{cost function} $c: \cX \times \cW \to \bR_+$. In this scenario, the adversary observes $y \in \cY$ and selects $w \in \cW$ to minimize $\bE [c(X,w) \mid Y=y]$. Essentially, this threat model is an adaptation of \cite{ACPS12milu} and \cite[Section II-B]{saeidian2023pointwise} where the adversary seeks to minimize a cost function instead of maximizing a gain function. To quantify the risk associated with such an adversary, we compare the smallest posterior expected cost with the smallest prior expected cost. Accordingly, we define 
\begin{equation}
\label{eq:cost_def}
    \Lambda_{c}(X \to y) \coloneqq \log \; \frac{\min_{w \in \mathcal W} \bE [c(X,w)]}{\min_{w \in \mathcal W} \bE [c(X,w) \mid Y=y]}. 
\end{equation}

Similar to the equivalence between the randomized function model and the gain function model established in \cite[Thm. 2]{saeidian2023pointwise}, we demonstrate below that \eqref{eq:U_def} and \eqref{eq:cost_def} provide equivalent characterizations of leakage.
\begin{theorem}
\label{thm:equivalence}
Let $y \in \cY$. For every randomized function of $X$, denoted by $U$, there exists a set $\cW_U$ and a cost function $c_{U} : \cX \times \cW_{U} \to \mathbb R_+$ such that $\Lambda_U(X \to y)  = \Lambda_{c_{U}} (X \to y)$. Conversely, for every cost function $c : \cX \times \cW \to \mathbb R_+$, there exists a randomized function of $X$, denoted by $U_c$, such that $\Lambda_c(X \to y) = \Lambda_{U_c}(X \to y)$.    
\end{theorem}

\begin{IEEEproof}
Fix $y \in \cY$. First, we argue that given a $U$ satisfying the Markov chain $U-X-Y$ there exists a non-negative cost function $c_U$ such that $\Lambda_{U}(X \to y) = \Lambda_{c}(X \to y)$. Given $U$, let the cost function $c_U: \cX \times \cU \to \bR_+$ be defined as $c_U(x,u) = 1 - P_{U \mid X=x}(u)$ with $x \in \cX$ and $u \in \cU$. Then, we have 
\begin{align*}
    \exp\Big( \Lambda_{c_U}(X \to y) \Big) &= \frac{\min_{u \in \mathcal \cU} \bE [c_U(X,u)]}{\min_{u \in \mathcal U} \bE [c_U(X,u) \mid Y=y]}\\[0.7em]
    &= \frac{\min_{u \in \cU} \sum_x c_U(x,u) P_X(x)}{\min_{u \in \cU} \sum_x c_U(x,u) P_{X \mid Y=y}(x)}\\[0.7em]
    &= \frac{\min_{u \in \cU} \sum_x (1 - P_{U \mid X=x}(u)) P_X(x)}{\min_{u \in \cU} \sum_x (1 - P_{U \mid X=x}(u)) P_{X \mid Y=y}(x)}\\[0.7em]
    &= \frac{1 - \max_{u \in \cU} \sum_x P_{U \mid X=x}(u) P_X(x)}{1 - \max_{u \in \cU} \sum_x P_{U \mid X=x}(u) P_{X \mid Y=y}(x)}\\[0.7em]
    &= \exp\Big( \Lambda_{U}(X \to y) \Big).
\end{align*}
Observe that the above construction works even when $\Lambda_{U}(X \to y) = \infty$ in which case we also get $\Lambda_{c_U}(X \to y) = \infty$.

Now, we show that for each non-negative cost function $c$, there exists $U$ satisfying the Markov chain $U-X-Y$ such that $\Lambda_{c}(X \to y) = \Lambda_{U}(X \to y)$. Fix a cost function $c$. Without loss of generality, we assume that $0 \leq c(x,w) \leq 1$ for all $x \in \cX$ and $w \in \cW$. This can be achieved through normalizing $c$ by $\max_{x,w} c(x,w)$. 

First, suppose $\Lambda_c(X \to y) < \infty$. Let $k$ be a large integer. We construct two randomized functions of $X$ denoted by $S$ and $T$ both on the same alphabet $[k+1]$. Let 
\begin{gather*}
    w_S \in \argmin_{w} \sum_x c(x, w) P_{X}(x),\\
    w_T \in \argmin_{w} \sum_x c(x,w) P_{X \mid Y=y}(x),
\end{gather*}
where $w_S$ denotes the adversary's optimal choice prior to observing $y$ and $w_T$ denotes the adversary's optimal choice after observing $y$. For all $x \in \cX$, let
\begin{gather*}
    P_{S \mid X=x}(i) = \frac{c(x,w_S)}{k}, \quad i \in [k],\\ 
    P_{S \mid X=x}(k+1) = 1 - c(x,w_S),\\
    P_{T \mid X=x}(i) = \frac{c(x,w_T)}{k}, \quad i \in [k],\\ 
    P_{T \mid X=x}(k+1) = 1 - c(x,w_T). 
\end{gather*}
Let $0 \leq \delta \leq 1$. We define $U_\delta$ as the mixture of $S$ and $T$ with parameter $\delta$, that is, 
\begin{equation*}
    U_\delta = \begin{cases}
        S & \text{with probability} \; \delta,\\
        T & \text{with probability} \; 1-\delta. 
    \end{cases}
\end{equation*}
Then, $P_{U_\delta \mid X=x}(i) = \delta P_{S \mid X=x}(i) + (1 - \delta) P_{T \mid X=x}(i)$ for $i \in [k+1]$ and we get 
\begin{align*}
    &\inf_{P_{\hat U}}\; \bP [U_\delta \neq \hat{U}] = \min_{u \in [k+1]} \sum_{x} (1-P_{U_\delta \mid X=x}(u)) P_X(x)\\[0.5em]
    &=\min \Bigg \{\sum_{x} \Big( 1 - \delta\frac{c(x,w_S)}{k} - (1 - \delta) \frac{c(x,w_T)}{k}\Big) P_X(x), \sum_{x} \Big(\delta c(x,w_S) + (1 - \delta) c(x,w_T) \Big) P_X(x) \Bigg\}\\[0.5em]
    &=\sum_{x} \Big(\delta c(x,w_S) + (1 - \delta) c(x,w_T) \Big) P_X(x), 
\end{align*}
where the last equality follows by letting $k \to \infty$. Similarly, we have 
\begin{align*}
    &\inf_{P_{\hat U \mid Y}} \; \bP [U_\delta \neq \hat{U} \mid Y=y] = \min_{u \in [k+1]} \sum_{x} (1-P_{U_\delta \mid X=x}(u)) P_{X \mid Y=y}(x)\\[0.5em]
    &=\min \Bigg \{\sum_{x} \Big( 1 - \delta\frac{c(x,w_S)}{k} - (1 - \delta) \frac{c(x,w_T)}{k}\Big) P_{X \mid Y=y}(x), \sum_{x} \Big(\delta c(x,w_S) + (1 - \delta) c(x,w_T)\Big) P_{X \mid Y=y}(x) \Bigg\}\\[0.5em]
    &= \sum_{x} \Big(\delta c(x,w_S) + (1 - \delta) c(x,w_T)\Big) P_{X \mid Y=y}(x), 
\end{align*}
and, once again, the last equality follows by letting $k \to \infty$. Therefore, we have
\begin{equation*}
    \exp \Big( \Lambda_{U_\delta} (X \to y) \Big) = \frac{\sum_{x} \Big(\delta c(x,w_S) + (1 - \delta) c(x,w_T) \Big) P_X(x)}{\sum_{x} \Big(\delta c(x,w_S) + (1 - \delta) c(x,w_T)\Big) P_{X \mid Y=y}(x)}.
\end{equation*}

Next, note that when $\delta = 1$ we have 
\begin{align*}
    \exp\Big(\Lambda_{U_1} (X \to y)\Big) &= \exp\Big(\Lambda_S(X \to y) \Big)\\
    &= \frac{\sum_x c(x,w_S) P_{X}(x)}{\sum_{x} c(x,w_S) P_{X \mid Y=y}(x)}\\[0.7em]
    &\leq \frac{\sum_x c(x,w_S) P_{X}(x)}{\sum_{x} c(x,w_T) P_{X \mid Y=y}(x)}\\
    &= \frac{\min_w \sum_x c(x, w) P_{X}(x)}{\min_w \sum_x c(x, w) P_{X \mid Y=y}(x)}\\[0.7em]
    &= \exp \Big(\Lambda_c(X \to y) \Big), 
\end{align*}
and when $\delta = 0$ we have 
\begin{align*}
    \exp\Big(\Lambda_{U_0} (X \to y)\Big) &= \exp\Big(\Lambda_T(X \to y) \Big)\\
    &= \frac{\sum_x c(x,w_T) P_{X}(x)}{\sum_{x} c(x,w_T) P_{X \mid Y=y}(x)}\\[0.7em]
    &\geq \frac{\sum_x c(x,w_S) P_{X}(x)}{\sum_{x} c(x,w_T) P_{X \mid Y=y}(x)}\\
    &= \frac{\min_w \sum_x c(x, w) P_{X}(x)}{\min_w \sum_x c(x, w) P_{X \mid Y=y}(x)}\\[0.7em]
    &= \exp \Big(\Lambda_c(X \to y) \Big).  
\end{align*}
In other words, we have $\Lambda_{U_1}(X \to y) \leq \Lambda_c(X \to y) \leq \Lambda_{U_0}(X \to y)$. Then, due to the continuity of the mapping $\delta \mapsto \Lambda_{U_\delta}(X \to y)$, there exists $\delta^* \in [0,1]$ such that $\Lambda_{U_{\delta^*}} (X \to y) = \Lambda_c(X \to y)$. 

Finally, we consider the case $\Lambda_c(X \to y) = \infty$. In this case, there exists a proper subset of $\supp(P_X)$ denoted by $\cE$ such that $P_{X \mid Y=y}(\cE) = 1$ and $c(x,w_T) = 0$ for all $x \in \cE$. Let $U$ be a binary random variable described by $P_{U \mid X=x}(0)  = 1$ for all $x \in \cE$ and $P_{U \mid X=x}(0) = \frac{1}{2}$ for $x \in \cX \setminus \cE$. Then, we have 
\begin{align*}
    \Lambda_U(X \to y) &= \log \frac{1 - \max \{P_U(0), P_U(1) \}}{1 - \max_u \sum_{x \in \cX} P_{U \mid X=x}(u) P_{X \mid Y=y}(x)}\\[0.7em]
    &=\log \frac{1 - \max \{\frac{1}{2} + \frac{P_X(\cE)}{2}, \frac{P_X(\cX \setminus \cE)}{2} \}}{1 - P_{X \mid Y=y}(\cE)}\\
    &= \infty, 
\end{align*}
since $0 < P_X(\cE) < 1$. 
\end{IEEEproof}

\begin{corollary}
By Theorem~\ref{thm:equivalence}, $\Lambda_{P_{XY}}(X \to y)$ can alternatively be defined as 
\begin{equation}
\label{eq:lambda_sup_c}
    \Lambda_{P_{XY}}(X \to y) \coloneqq \sup_{c} \; \Lambda_c(X \to y),  
\end{equation}
where the supremum is over all non-negative cost functions $c$. 
\end{corollary}

\begin{remark}[Closedness under pre-processing]
\label{rem:pre_proc}
An important property of $\Lambda_{P_{XY}}(X \to y)$ apparent from \eqref{eq:lambda_sup_U} is that it is \emph{closed under pre-processing}, that is, $\Lambda_{P_{UY}} (U \to y) \leq \Lambda_{P_{XY}} (X \to y)$ for all $U$'s satisfying the Markov chain $U-X-Y$ and all $y \in \cY$. Closedness under pre-processing implies that the amount of information leaking about functions of $X$ can never exceed the amount of information leaking about $X$ itself.  
\end{remark}

\subsection{Relationship to Maximal Cost Leakage and Maximal Realizable Cost}
\label{ssec:lower_bound_discuss}

In \cite{IssaMaxL}, two notions of information leakage are defined also by considering risk-averse adversaries. Specifically, \citet{IssaMaxL} define \emph{maximal cost leakage} as 
\begin{equation}
\label{eq:mcl}
    \cL^{\sfc}(X \to Y) \coloneqq \sup_{\substack{U: U-X-Y, \\ \hat{\cU}, c:\cU \times \hat{\cU} \to \bR_+}} \; \log \; \frac{\inf_{\hat u \in \hat{\cU}} \bE[c(U,\hat u)]}{\inf_{\hat u(\cdot)} \bE[c(U, \hat u(Y))]}, 
\end{equation}
and \emph{maximal realizable cost} as 
\begin{equation}
\label{eq:mrc}
    \cL^{\sfr \sfc}(X \to Y) \coloneqq \sup_{\substack{U: U-X-Y, \\ \hat{\cU}, c:\cU \times \hat{\cU} \to \bR_+}} \; \log \; \frac{\inf_{\hat u \in \hat{\cU}} \bE[c(U,\hat u)]}{\min_{y \in \cY}  \inf_{\hat u \in \hat{\cU}} \bE[c(U, \hat u) \mid Y=y]}.  
\end{equation}
Note that \eqref{eq:mcl} and \eqref{eq:mrc} measure the costs associated with guessing randomized functions of $X$, whereas \eqref{eq:cost_def} measures the cost associated with guessing $X$ itself. 

At first glance, maximal cost leakage appears to capture the expected value of the leakage over the outcomes of $Y$. However, according to  \eqref{eq:cost_def} and \eqref{eq:lambda_sup_c}, maximal realizable cost can be expressed as 
\begin{align*}
\cL^{\sfc}(X \to Y) &= \sup_{U : U-X-Y} \left(-\log \; \bE_{Y \sim P_Y} \left[\exp \Big(-\Lambda_{P_{UY}} (U \to Y) \Big) \right] \right)\\
&\leq \sup_{U : U-X-Y} \bE_{Y \sim P_Y} \left[\Lambda_{P_{UY}} (U \to Y) \right]\\
&\leq \bE_{Y \sim P_Y} \left[\sup_{U : U-X-Y} \Lambda_{P_{UY}} (U \to Y) \right]\\  
&= \bE_{Y \sim P_Y} \left[\Lambda_{P_{XY}} (X \to Y) \right],  
\end{align*}
where the first inequality is due to Jensen's inequality and the last equality follows from Remark~\ref{rem:pre_proc}. Note that due to the strict concavity of $\log(\cdot)$ Jensen's inequality is strict. Consequently, maximal cost leakage does not describe the expected value of the information leakage but underestimates it. In addition, maximal realizable cost can be expressed as 
\begin{align*}
    \cL^{\sfr \sfc}(X \to Y) &= \max_{y \in \cY} \; \sup_{U:U-X-Y} \; \Lambda_{P_{UY}} (U \to y)\\
    &= \max_{y \in \cY} \; \Lambda_{P_{XY}} (X \to y),
\end{align*}
implying that the supremum over $U$'s is superfluous in \eqref{eq:mrc}.

More generally, defining the leakage for each $y \in \cY$ has the advantage that we can view $\Lambda_{P_{XY}} (X \to Y)$ as a random variable that can be restricted in various ways. Hence, we are not restricted to calculating the average or maximum leakage but we may also restrict the tail of $\Lambda_{P_{XY}} (X \to Y)$ or its different moments. 

\subsection{Discussion}

According to Theorem \ref{thrm:operationalization}, a lower bound on the information density can be equivalently formulated as an upper bound on the quantity
\begin{equation}
\label{eq:remInvInfDens}
    \max_{x\in\mathcal X} \log \frac{P_X(x)}{P_{X|Y=y}(x)}.
\end{equation}
It is interesting to note that this prior-to-posterior ratio does \emph{not} fulfill an axiomatic requirement for information leakage measures in the \emph{quantitative information flow} (QIF) framework. In particular, \citet{smith2009foundations} posited that a measure of information should describe the difference between our initial uncertainty about a random variable and our remaining uncertainty after observing a correlated quantity. In other words, any leakage measure should satisfy   
\begin{equation}
\label{eq:info_leakage}
    \text{information leakage} = \text{initial uncertainty} - \text{remaining uncertainty}.
\end{equation}
Now, suppose $X$ is a binary random variable with $P_X(0) = 0.99$ and posterior distribution $P_{X \mid Y=y}(0) = 1$. Then, $\Lambda_{P_{XY}}(X \to y) = \infty$ while the initial uncertainty in $X$ is finite according to all the usual measures of uncertainty, e.g., the Rényi entropies~\cite{renyi1961entropy}. As a result, an upper bound on \eqref{eq:remInvInfDens} cannot be interpreted as limiting a form of information leakage in the sense of \eqref{eq:info_leakage}.  


Nonetheless, Theorem \ref{thrm:operationalization} together with the inferential statements in \cite{saeidian2023inferential,grosse2023extremal} shows that imposing both and upper \emph{and} a lower bound on $i(x;y)$ ensures that in addition to not being able to infer the \emph{correct} value of the private realization (upper bound), an adversary can also not say with too much certainty which value of the private variable was \emph{not} realized (lower bound). Constraining an adversary's inference ability in this way is directly related to the concept of \emph{absolute disclosure prevention} in the PML framework \cite{saeidian2023inferential}. Absolute disclosure prevention can also be ensured by picking $\epsilon$ in the PML high-privacy regime \cite{grosse2023extremal}. Therefore, if an $\epsilon$-PML guarantee is chosen such that $\epsilon$ is in this high-privacy regime, then the guarantees of PML and (A)LIP become very similar and the difference between the two measures is reduced to an explicit versus an implicit lower bound on $i(x;y)$.

\section{Conclusion}
\label{sec:conclusion}
In this paper, we provided an explicit relation between upper and lower bounds on information density by exploiting row-stochasticity conditions on the transitioning kernels. We utilized these bounds to derive new relationships between (asymmetric) local information privacy, pointwise maximal leakage and local differential privacy and provided an application of these relations to optimal privacy mechanism design. We further showed that a lower bound on information density can be operationalized as an adversary aiming to minimize the probability of incorrectly guessing the value of the private realization. These results close the gap between the definitions of LIP, ALIP and PML, and make all three measures and their relation to each other understandable. It is then left to the user to decide which operational meaning is relevant for a specific privacy problem at hand, and set upper and lower bounds accordingly. 


\bibliographystyle{IEEEtranN}
\footnotesize
\balance
\bibliography{bibtex/bib/IEEEabrv, bibtex/bib/thisbib}

\end{document}